\documentclass[aps,pre,twocolumn,amssymb,showpacs,groupedaddress]{revtex4-2}
\usepackage{graphicx,color}
\usepackage[normalem]{ulem}
\usepackage{amsmath}
\usepackage{tabularx}
\def\beq{\begin{equation}}
\def\eeq{\end{equation}}
\def\bea{\begin{eqnarray}}
\def\eea{\end{eqnarray}}

\def\cL{\mathcal{L}}
\def\LH{L_B}
\def\LT{L_T}
\def\cR{R}
\def\cLH{\mathcal{L}_B}
\def\cLT{\mathcal{L}_T}
\def\Wnh{\mathcal{W}}
\def\dq{\Delta q}
\def\dw{\Delta w}
\def\q0{q_0}

\begin{document}
\definecolor{colortodo}{RGB}{255,0,0}
\newcommand{\red}[1]{{\color{colortodo}#1}}

\title{Energetics of twisted elastic filament pairs}

\author{Julien Chopin,$^{1}$ Animesh Biswas$^{2}$ and Arshad Kudrolli$^{2}$}
\affiliation{$^{1}$ Instituto de F\'isica,  Universidade Federal da Bahia, Salvador-BA 40170-115, Brazil\\$^{2}$Department of Physics, Clark University, Worcester, Massachusetts 01610, USA}
\date{\today}

\begin{abstract}
We investigate the elastic energy stored in a filament pair as a function of applied twist by measuring torque under prescribed end-to-end separation conditions. We show that the torque increases rapidly to a peak with applied twist when the filaments are initially separate, then decreases to a minimum as the filaments cross and come into contact. The torque then increases again while the filaments form a double helix with increasing twist. A nonlinear elasto-geometric model that combines the effect of geometrical nonlinearities with large stretching and self-twist is shown to capture the evolution of the helical geometry, the torque profile, and the stored energy with twist. We find that a large fraction of the total energy is stored in stretching the filaments, which increases with separation distance and applied tension. We find that only a small fraction of energy is stored in the form of bending energy, and that the contribution due to contact energy is negligible. Further, we provide analytical formulas for the torque observed as a function of the applied twist and the inverse relation of the observed angle for a given applied torque in the Hookean limit. Our study highlights the consequences of stretchablility on filament twisting which is a fundamental topological transformation relevant to making ropes, tying shoelaces, actuating robots, and the physical properties of entangled polymers. 
\end{abstract}

\maketitle
\section{Introduction}

\begin{figure}
\includegraphics[width=8.5cm]{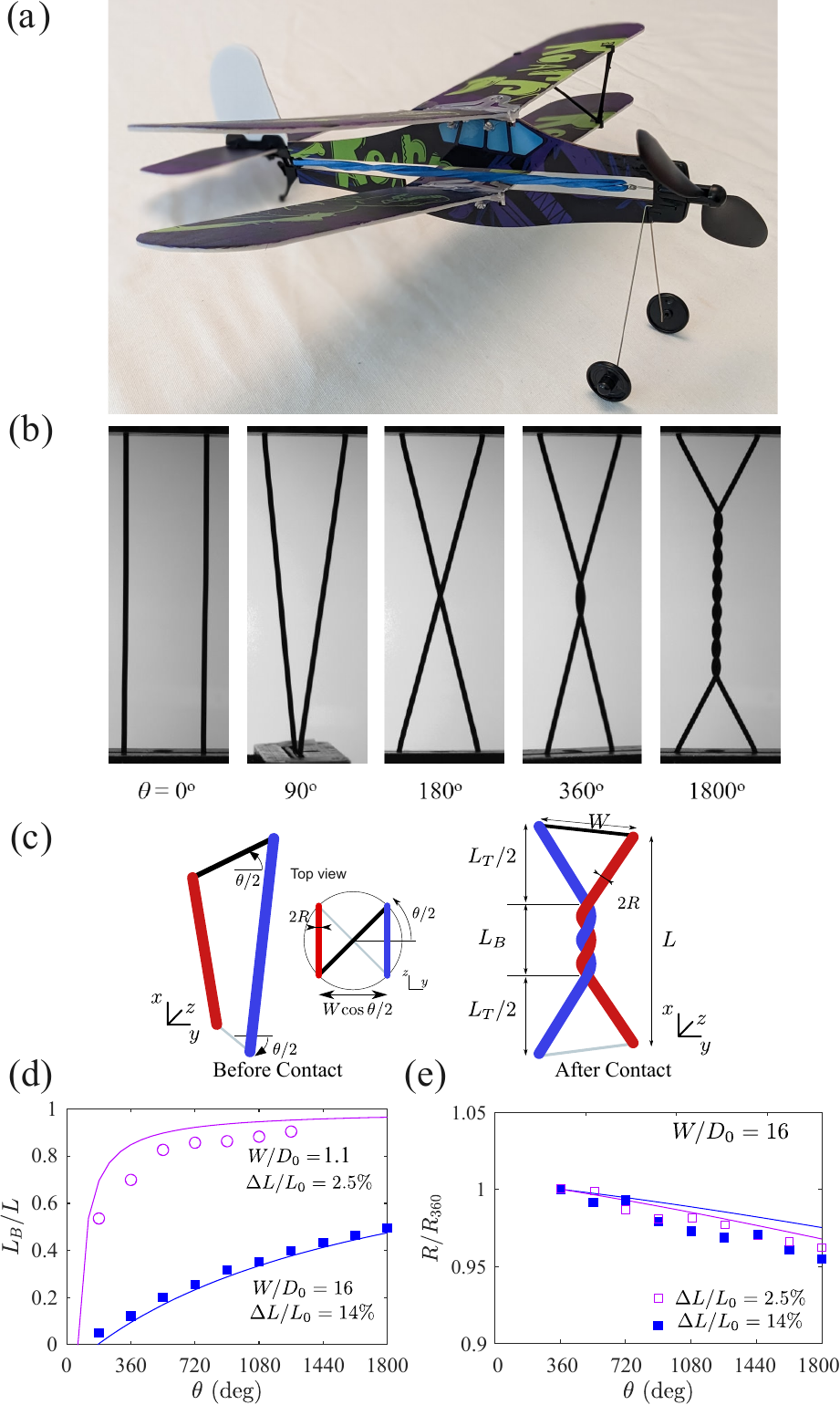}
	\caption{(a) A rubber band airplane powered by elastic energy. (b) Images of the filaments twisted about the central axis while clamped at a fixed distance $L/L_0 = 1.14$ and $W/D_0=16$ apart. (c) Schematic of the filament geometry before the filaments come in contact and after contact when a helical bundle 
 forms at the center. (d) The measured evolution of the bundle size $L_B/L$ for various $W$ and comparison with Eq.~(\ref{eq:LB}). (e) The bundle radius as a function of twist normalized by the radius $R_{360}$ at $\theta = 360^{\circ}$ and Eq.~(\ref{eq:R}), where $\lambda$ is obtained using the elasto-geometric model.
  } 
	\label{fig:intro}
\end{figure}

Twisting and braiding filaments are relevant to making ropes, tangling, and are an important construct in biomaterials~\cite{Bohr2011,legrain2016let,weiner2020mechanics,Slepukhin2021,Gao2021,Plumb2022,seguin2022twist,patil2023}. It is an important step in tying a knot, where the contact structure for filaments meeting even at $90^o$ can be quite complex~\cite{Grandgeorge2021}. A common element in windup toys (see Fig.~\ref{fig:intro}(a)), elastomeric filaments are being developed towards applications in novel energy storage and rapid mechanical actuation~\cite{Moshe2004,Guzek2012,kwon2014high,Haines2016}. The twist of a single filament itself can be quite subtle~\cite{charles2019topology}. While the filament deforms uniformly when twist is applied at its ends about its central axis, a helical instability occurs above a critical twist depending on the applied tension, which causes the central line of the filament to follow a helical path about the central axis~\cite{Thompson1996,Audoly2016}. The pitch cannot be less than the diameter of the filament limiting the phase space of possible winding angles that a filament can have~\cite{Olsen2012,Grason2015}. Although polymer filaments are typically hyperelastic, their analysis has often treated them as being inextensible~\cite{Neukirch2002,Grason2012,Panaitescu2018}. The effective radius and contraction of filament bundles have been measured as a function of applied twist and shown to deviate from models based on volume conservation~\cite{Hanlan2023}. Braiding and torque of extended and twisted DNA pairs was investigated by Charvin, et al.~\cite{Charvin2005} with experiments and models which mainly considered bending energy.   

Here, we investigate the energy stored in a pair of elastic filaments by measuring and analyzing the torque as a function of applied twist with an elasto-geometric model. This system is not only important in its own right, but also forms a basis to understand the topology and energetics of multi-filament bundles where geometric frustration leads to further complexity in arriving at their configurations with twist~\cite{Bruss2013,Panaitescu2018}.  A priori one may expect energy contributions from bending, stretching, and twisting of each filament, and the filament-filament contact. We consider a regime where the material deformations themselves can be considered as linear or neo-Hookean, but where the geometry of the configurations leads to a complex response. We show that energy is essentially stored in twisting and stretching the filaments which vary in relative magnitude depending on the system geometry and applied twist. The contribution of bending and inter-filament contact are found to be negligible over the observed twist angles. 

In the following, we first discuss the experimental measurements of the geometry and the torque encountered as a function of the applied twist with elastic filament-pairs in Section~\ref{sec:expt}. Then, we develop an elasto-geometric model by considering the energy stored assuming neo-Hookean response under large deformations  and compare it with the measured data in Section~\ref{sec:elastomodel}, and calculate the energy stored within stretch, bend, twist and contact components in Section~\ref{sec:comp}.  Before concluding, we provide analytical formulas in the limit of Hookean response, and discuss the twist angle reached under prescribed torque conditions in Section~\ref{sec:anal}. 

\section{Experiments}
\label{sec:expt}
We consider two elastic filaments with radius $R_0$ (diameter $D_0 = 2R_0$) and relaxed length $L_0$, that are initially parallel and separated by a distance $W$ and clamped at their ends as shown in Fig.~\ref{fig:intro}(b). The filaments can be also pre-stretched by a distance $\Delta L$, resulting in a length $L = L_0 + \Delta L$. The filaments are twisted through end-to-end angle $\theta$ about the central axis which parallels the initial orientation of the filaments.  The filaments are composed of silicone (USA Industrials) and have a circular cross section of radius $R_0 = 1.27$\,mm. We perform measurements of torque with a Mark-10 MR50-10Z torque sensor fixed to the top clamp.  The bottom clamp is rotated through a prescribed angle $\theta$ with a computer controlled microstepper motor, and images are taken with a 9.6\,megapixel Pixelink PL-D729MU camera while $\theta$ is incremented in $1^{\circ}$\,steps. 

Figure~\ref{fig:intro}(b) shows a sequence of images as the filaments are twisted around each other~5 times when $L=17.2$\,cm and $W=4$\,cm. As shown schematically in Fig.~\ref{fig:intro}(c), the filaments can be represented as straight lines before contact at twist angle $\theta_c$ if the curvature due to bending at the clamps is negligible, and then as a double helix connected to essentially straight filament segments after contact. We measure and plot the helical bundle length $L_B$ in Fig.~\ref{fig:intro}(d), and the helical bundle radius $R$ normalized by the radius $R_{360}$ after one full twist corresponding to $\theta = 360^{\circ}$ in Fig.~\ref{fig:intro}(e). (As a consequence of geometry, this radius can be seen to be the same as the radius of the filament if the contact deformation is negligible.) We observe that the double helix grows in length while its radius decreases slowly with increasing $\theta$ depending on $W$.  

\begin{figure*}
\includegraphics[width=18cm]{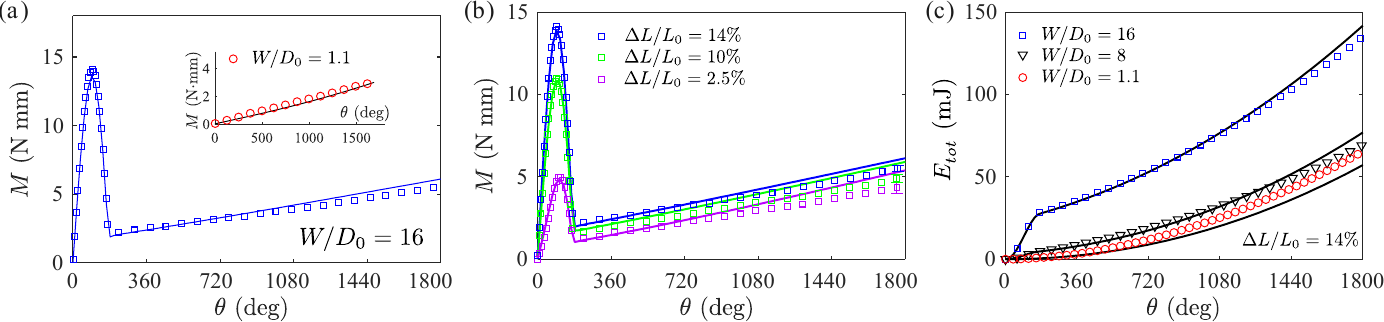}
	\caption{(a) The measured (marker) torque versus twist angle. Comparison with 
 Neo-Hookean model (thin blue/gray line). Inset: The measured torque when $W = 2R_0$ compared with model. (b) Measured and calculated torque versus twist angle for 2-filament twist for 
 $W =4$\,cm for various pre-stretch conditions $L/L_0$. The peak before filaments come in contact increases with $L$. The calculated torque using stretching, twist and bending energy assuming Neo-Hookean model with no fit parameters. Good overall agreement is observed. The measurement errors are indicated by bars on the last set of markers. (c) The  measured (markers) and corresponding calculated (lines with corresponding color/shade) energy as a function of twist for various $W$. The stored energy increases with separation distance.} 
	\label{fig:torque}
\end{figure*}

These changes in the geometry can be seen to have significant effects on the torque required to twist the filaments. Figure~\ref{fig:torque}(a) shows the torque $M$ measured as a function of $\theta$ corresponding to $L/W = 4.2$ and $W/D_0 = 16$. We observe that the torque increases rapidly before decreasing to a minimum when the filaments come in contact, and then rises again.  The appearance of a peak may seem surprising, but it has been seen previously in twisted DNA~\cite{Charvin2005}. A similar feature has been also noted in a twisted sheet, besides further non-monotonous behavior at higher twists due to shape transformations~\cite{Chopin2022}. When $W \approx D_0$, the peak is absent and a monotonic increasing torque is observed (see Inset to Fig.~\ref{fig:torque}(a)).  As illustrated in Fig.~\ref{fig:torque}(b), decreasing pre-stretch is observed to lead to an overall decrease in $M$, but the peak is present even when $L \simeq L_0$, and the overall variation of $M$ with $\theta$ is similar for all the cases. 

The elastic energy stored in the filaments can be obtained by integrating the torque over the twist angle, i.e. $E_{tot} = \sum_0^\theta M \Delta\theta$, where $\Delta\theta$ corresponds to $1^{\circ}$ in our experiments. The corresponding data are plotted in Fig.~\ref{fig:torque}(c) and show that $E_{tot}$ increases monotonically with $\theta$. The energy is a smooth function without kink or discontinuity at the point of contact as it is obtained by integrating the torque which is a continuous function of the twist angle. Further, it can be observed that the energy increases rapidly with increasing $W$ before filaments come in contact, and then increases more slowly after they cross. Thus, the total stored energy can be increased significantly by simply separating the filaments for the same amount of elastic material.  

\section{Elasto-geometric Model}
\label{sec:elastomodel}
\begin{figure}
    \centering
    \includegraphics[width=8.5cm]{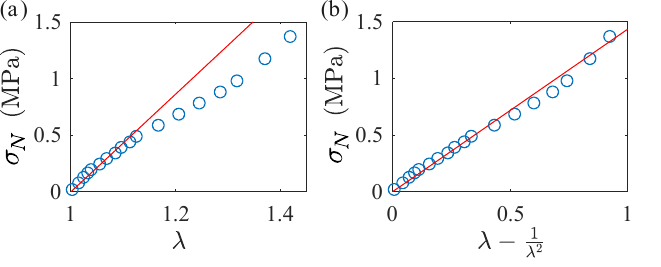}
     \caption{(a) Nominal stress $\sigma_N$ as a function of $\lambda$ is sublinear for large enough extensions. (b) $\sigma_N$ scales linearly as a function of $\lambda - \frac{1}{\lambda^2}$,, where $\lambda = L/L_0$ since $\cL = L$ as the filaments are not twisted during these measurements. The linear fit confirms the neo-Hookean nature of the filament, and yields shear modulus $\mu = 1.43 \pm 0.1$\,MPa. 
    }
    \label{fig:ss}
\end{figure}

We now develop a model which can capture the geometry of the twisted filament, the torque, and the stored elastic energy as a function of twist angle. An established method to understand such problems is to consider the elastic energy stored in the system, and consider its variation with appropriate stretch parameter to obtain the force or torque required as a function of applied boundary conditions~\cite{Audoly2010}.

Figure~\ref{fig:ss}(a) shows the applied nominal stress $\sigma_N = F/S_0$, where $F$ is the applied longitudinal force and $S_0 = (\pi/4)D_0^2$, versus measured stretch $\lambda = L/L_0$ along the length of the filaments. We observe that the stress increases linearly initially, but then becomes weakly nonlinear. This is a hallmark of a neo-Hookean response. Then, the strain-energy function~\cite{Ogden1984}:
\bea
\Wnh = \frac{1}{2}\mu\left(\lambda_1^2+\lambda_2^2+\lambda_3^2-3 \right),\quad \lambda_1\lambda_2\lambda_3=1,
\label{eq:energy_gal}
\eea
where $\lambda_i$ are the principal stretches and $\mu$ is the shear modulus. For uniaxial stretching, $\lambda_1=\lambda$ along the filament axis, $\lambda_2 = \lambda_3 = 1/\sqrt{\lambda}$. Hence, the radius $R$ of a stretched filament is reduced due to volume conservation and is given by: 
\bea
\cR=R_0/\sqrt{\lambda}. 
\label{eq:R}
\eea 
Accordingly, for a uniaxially stretched filament, the nominal stress $\sigma_N$ is given by~\cite{Ogden1984},
\bea
\sigma_N = \mu \left( \lambda - \frac{1}{\lambda^2} \right).
\label{eq:sig_N}
\eea

In Fig.~\ref{fig:ss}(b), we find that the data agrees with the neo-Hookean model with $\mu = 1.43 \pm 0.1$\,MPa over the typical stretches experienced in our system. 

We parameterize the kinematics of the filament pairs to obtain the stretch as a function of twist. Before contact ($\theta < \theta_c$), the shapes of the filaments are modeled by straight lines connecting the top and bottom clamps (see Fig.~\ref{fig:intro}(c)). One may expect a region of order $R_0$ from the clamp to be curved, but we neglect this as being relatively small, as can be noted from Fig.~\ref{fig:intro}(b). After contact ($\theta \geq \theta_c$), as seen from Fig.~\ref{fig:intro}(b), the filament geometry can be divided into a central helicoidal section of length $\LH$, and two triangular sections next to the clamps of height $L_T$ in total, with $L = \LH+L_T$. The central shape where the filaments are in enduring contact is modeled as a double helix of length $\LH$, radius $\cR$, and helix angle $\beta$. Therefore, the curvature in the helix is given by~\cite{o2006elementary}:
\bea
\kappa = \frac{1}{R} \frac{\tan^2 \beta}{1+\tan^2 \beta},
\label{eq:kappa}
\eea
with $\tan \beta = R (\theta-\theta_c)/L_B = \dq/l_B$, $\dq = R (\theta - \theta_c)/L$ and $l_B = L_B/L$.
The torsion $\tau$ is assumed to be homogeneous along the entire filament and given by $\tau = \theta / \cL$ where $\cL$ is the curvilinear length of a filament.

As can be seen from the top view in Fig.~\ref{fig:intro}(c), the distance between the filaments in the $yz$-plane is $W\cos \theta/2$. Thus, contact occurs when $W\cos \theta_c/2=2R$, yielding
\bea
\theta_c = 2 \arccos\left(\frac{2R}{W} \right)
\label{eq:thetac}
\eea
for $R/W < 1/2$. If the filaments are initially close to each other, i.e. $W \approx D_0$, $\theta_c$ varies rapidly with pre-stretch and is given by $\theta_c \approx 2 \sqrt{\lambda-1}$. 

After parametrizing the shape, we can evaluate the filament deformation using the longitudinal stretch factor
\bea
\lambda = \cL/L_0.
\label{eq:lambda}
\eea

Before contact, as the filaments are modeled by straight lines, $\cL$ is given as the distance between the top and bottom clamps, 
hence:
\bea
\lambda = \lambda_0 \sqrt{1 + \left(\frac{W}{L} \sin \frac{\theta}{2} \right)^2}\,,\,\theta<\theta_c,
\label{eq:lambda1}
\eea
where $\lambda_0 = L/L_0$ is the stretch factor before twist where $\theta = 0$. 

After contact, the inserted twist in the helix is $\Delta \theta = \theta-\theta_c$ and the arc length of the helix is given by
\bea
\cLH = L\sqrt{l_B^2 + \dq ^2},
\label{eq:LBs}
\eea
with $\dq = \Delta \theta R /L$, and $l_B = \LH/L$. Outside the helical region, the filaments are not in contact and are assumed to be straight lines. The filaments form two triangles of height $\LT/2 = (L-\LH)/2$ each and base size $W$ in that region. We can calculate the triangle side length $\cLT$ by evaluating the distance between the clamped end of one of the filaments, say the right one, and the point at which the filament enters into the helix at $(\LH/2,\cR \cos(\Delta \theta/2),\cR \sin(\Delta \theta/2)$. Thus, $(\cL_T/2)^2 = (\LT/2)^2 + (W/2\cos(\theta/2)-\cR\cos (\Delta \theta/2))^2 + (W/2 \sin(\theta/2) - \cR \sin (\Delta \theta/2))^2$. Using Eq.~(\ref{eq:thetac}) and trigonometric identities, we have
\bea
\cLT = L\sqrt{l_T^2 + \dw ^2},
\label{eq:LTs}
\eea
where $l_T = \LT/L$, and $\dw = \sqrt{W^2-4\cR^2}/L$. 

From the total arc length of the filament given by $\cL = \cLH + \cLT$ and Eq.~(\ref{eq:lambda}), we obtain 
\bea
\lambda = \lambda_0\sqrt{l_T^2+\dw^2} +\lambda_0\sqrt{l_B^2 +\dq^2}\,,\,\theta\geq \theta_c.
\label{eq:lambdahelix}
\eea
Thus, $\lambda$ is completely known once $l_T$ and $l_B$ are determined. Since $1=l_B+l_T$,  we only need to determine, say, $l_B$ which can be found by minimizing the total elastic energy $E_{tot}$. 

In the two-filament system, the total energy can be written as 
\begin{equation}
E_{tot} =  E_{t} + E_{s} + E_{b} + E_{c},
\label{eq:Etot}
\end{equation}
where  $E_t$ is the energy associated with twisting the filaments about their axis, $E_s$ is the energy associated with stretching the filaments, $E_b$ is the energy associated with bending the filaments into the helical shapes that they adopt with twist, and $E_c$ is the energy due to filaments contact contractions. The twist energy of the two filaments is given by~\cite{rivlin1951large}, 
\bea
E_t  =   \frac{1}{2}\mu V \frac{\q0^2 }{\lambda},
\label{eq:Et}
\eea
where $q_0 = R_0 \theta/L_0$. $V = \pi R_0^2 L_0$ is the volume in the reference (zero-load) configuration. The twist-induced stretching energy is given by~\cite{rivlin1948large} 
\bea
E_s = \mu V \left(\lambda^2 +\frac{2}{\lambda} -3 \right)-E_s^{\circ},
\label{eq:Es}
\eea
where $E_s^{\circ} = \mu V \left(\lambda_0^2 +\frac{2}{\lambda_0} -3 \right)$ is the stretching energy after pre-stretching. With this definition, the total energy is zero after pre-stretching, consistent with our definition $E_{tot} = \sum_0^{\theta} M\Delta \theta$. The bending energy is given by~\cite{Ghatak2005}, $E_b = B/2\, (\cLH/\lambda)\, \kappa^2$, where $B = 3/4\, \mu \,\pi R_0^4 = 3/4\, \mu V\, R_0^2 / L_0$ is the bending rigidity. Thus,
\bea
E_b  = \frac{3}{8}\mu V \, \lambda(\kappa R)^2 \, \frac{\cLH}{\cL},
\label{eq:Eb}
\eea
 where we use the fact that $R_0^2/L_0 = \lambda^2 R^2/\cL$. The $\lambda$-dependence of $R$ is given by Eq.~(\ref{eq:R}). The inter-filament contact energy $E_c$ can be estimated by assuming the curvature is small enough that the curved contact can considered as a straight line. The Hertz contact energy of two parallel cylinders pressed against each other, forming a contact over a length $\cLH$, is given by $E_c = \pi/2\, \mu (\delta R)^2 \,\cLH$ \cite{johnson1987contact,Panaitescu2018}, where $\delta R$ is the indentation  of the two filaments upon self-contact. Therefore, we have:
\bea
E_c &=& \frac{1}{2}\mu V \left( \frac{\delta R}{R} \right)^2 \frac{\cLH}{\cL}\,.
\label{eq:Ec}
\eea
To obtain $\delta R$, we consider a force equilibrium between the normal inward force $(\sigma_N\, \pi R_0^2)\, \kappa\,  \cLH = (\sigma_N V)\, \kappa \,\lambda (\cLH/\cL)$ and the Hertz resisting force~\cite{johnson1987contact}, $d E_c/d( \delta R) = \mu V (\delta R/R^2) (\cLH/\cL)$, which gives:
\bea
\frac{\delta R}{R} \approx \frac{\sigma_N\,\lambda}{\mu}\, \kappa R.
\eea 
After substituting in Eq.~(\ref{eq:Ec}), this yields 
\bea
E_c
= \frac{1}{2}~\,\mu V \left(\lambda^2-\frac{1}{\lambda}\right)^2(\kappa R)^2\, \frac{\cLH}{\cL},
\eea
where we use $\sigma_N \lambda = \mu(\lambda^2-1/\lambda)$, following Eq.~(\ref{eq:sig_N}).

Then, we find numerically the bundle length $l_B$ which minimizes $E_{tot}$ and compare its evolution as a function of $\theta$ for various $W$ with experiments in Fig.~\ref{fig:intro}(d). We obtain good agreement without fitting parameters. Further, $R$ as a function of $\theta$ is calculated using Eq.~(\ref{eq:R}) and found to also be in overall good agreement with the measured values (see Fig.~\ref{fig:intro}(e)). Thus, the geometry of the twisted filaments can be captured by the elasto-geometric model by simply using energy minimization.  

Having found $l_B$, we evaluate $E_{tot}$, and plot it as a function of $\theta$ for the three different $W$ in Fig.~\ref{fig:torque}(c). We observe that it follows the measured values without any fitting parameters within experimental error. Finally, we obtain the torque profile by numerical differentiation since $M = \partial E_{tot}/\partial \theta$ to directly compare with the measurements. Figure~\ref{fig:torque}(a)  shows the comparison for $W=4$\,cm and for $W=D_0$ in the Inset to Fig.~\ref{fig:torque}(a), where quantitative agreement can be observed in each case without fitting parameters. Thus, we have a fully tested model of the geometry, torque and elastic energy stored in twisted two-filament pairs with neo-Hookean material properties.  

\section{Energy components} \label{sec:comp}
\begin{figure*}
    \centering
    \includegraphics[width=17cm]{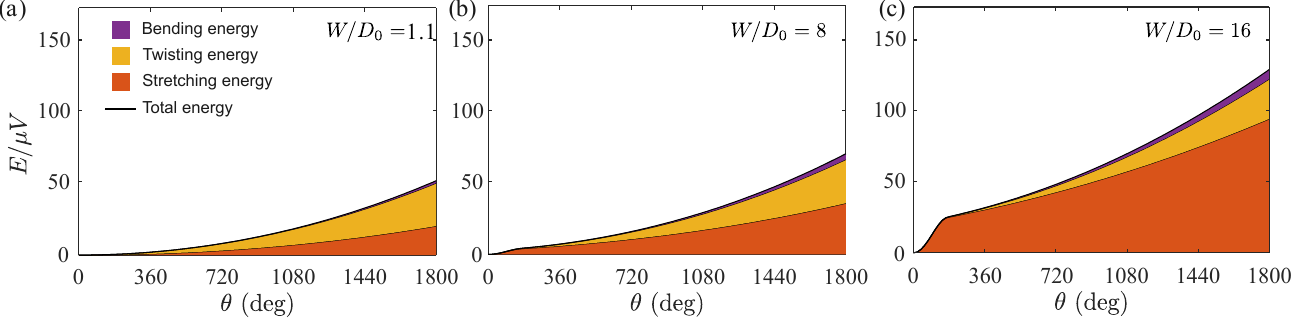}
     \caption{(a-c) The contribution of twisting, bending, and stretching energies to total stored energy as a function of twist for various $W$. The contact energy contribution is negligible and thus not plotted. ($\Delta L/L_0 = 0.14$, same as in Fig.~2(c)). 
    }
    \label{fig:Efrac}
\end{figure*}

We examine the various energy components to understand their relative contributions due to geometry and pre-stress. Figure~\ref{fig:Efrac}(a-c) shows $E_{tot}$ for the three different $W$. We observe that the contribution of twisting energy to total energy is larger than stretching when the filaments are close together, but stretching makes an increasingly greater contribution with increasing $W$. Furthermore, it can be noted that bending contributes a little to the total energy, and that the contribution of contact energy is in fact smaller than the line width, so we have not plotted it. 

\begin{figure}
    \centering
       \includegraphics[width=8.5cm]{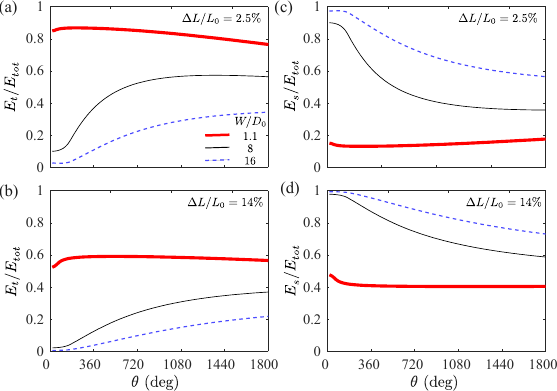}
    \caption{Energy fraction for various $W$ and $\lambda$ stored in twist (a,b), and in stretching (c,d). While most of the total energy is in the form of twist at lowest $\Delta L/L_0$ and $W$, an increasing fraction is stored in stretching the filaments with twist. Energy stored in stretching the filaments dominates as $W$ and $\Delta L/L_0$ are increased.}
    \label{fig:Efractions}
\end{figure}

To tease out the twist and stretching energy contributions to the total energy further, we plot the ratio $E_t/E_{tot}$ and $E_s/E_{tot}$ in Fig.~\ref{fig:Efractions} for the various $W$ and $\Delta L/L_0$. We observe that a large fraction of the total energy is stored in the twist mode when pre-stretch $\Delta L/L_0$ is small, but an increasing fraction is in the stretch mode as the filaments are stretched about the axis of rotation. 
In other words, increasing $W$ and $\Delta L$ for the same length of filament $L_0$ leads to increasing $E_{tot}$, with increasing fraction stored as stretching energy. This suggests a maximization strategy for total stored energy per unit filament length, by increasing pre-stretch and separation distance between twisted filaments one increase store energy for the same amount of material and number of twist turns. 

\section{Analytical formulas for energy and torque}\label{sec:anal}
We next examine some analytical limits of our elasto-geometric model in order to provide formulas of energy and torque in the case of twisted elastic filaments. 

\subsection{Filament pairs in contact in the Hookean limit}
When the two filaments are initially in contact, the calculation of $\lambda$ is straightforward as they are modeled by stretched helices, yielding $\lambda = \lambda_0 \sqrt{1 + (R\theta/L)^2}$ which corresponds to Eq.~(\ref{eq:lambdahelix}) with $L = \LH$ and $\Delta w = l_T = 0$. In the Hookean limit, for $\Delta L/L_0 \sim (R_0/L_0)^2 \ll 1$, we have $ \lambda \approx  \lambda_0 + \frac{1}{2}\left(R_0/L_0\right)^2 \theta^2$ and $\lambda_0 \approx 1+\Delta L/L_0 $. Thus, 
\bea
\frac{M}{\mu V} = \left(\frac{R_0}{L_0}\right)^2 \left(1 + 6 \frac{\Delta L}{L_0} \right)\theta.
\label{eq:M2R}
\eea
Therefore, the stretching of the filament introduces an additional term in the torque which scales in the Hookean limit as $(R_0/L_0)^2(\Delta L/L_0)$. 

\begin{figure}
    \centering
    \includegraphics[width=7cm]{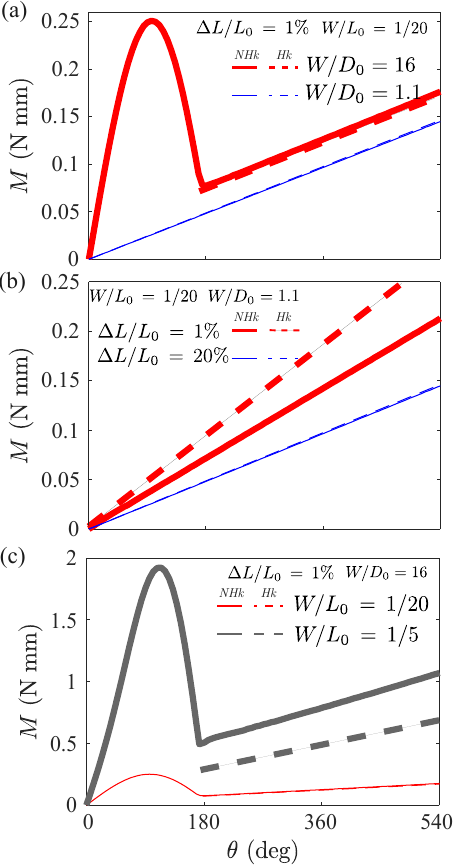}
    \caption{Torque as a function $\theta$ after contact 
    calculated from the Neo-Hookean (NHk) model (solid thick red and thin blue lines) and the Hookean (Hk) limit given by Eq.~(\ref{eq:torque_analin_aftercontact}) (dashed thick red and thin blue lines). (a) The Neo-Hookean and Hookean curves agree for small $\Delta L/L_0$. (b,c) The two estimates agree for sufficiently small $\Delta L/L_0$ for small enough $L/W$ (b), but disagree for large $W/L_0 = 1/5$ (c). 
    }
    \label{fig:linl}
\end{figure}

We compare the torques obtained using these analytical calculations assuming Hookean-response and those assuming neo-Hookean response in Fig.~\ref{fig:linl}(a). We observe that these calculations agree for sufficiently small $\Delta L/L_0$, but deviations are observed as $\Delta L/L_0$ is increased to 20\%.

\subsection{Filament pairs with separation}

\subsubsection{Before Contact}
Since the filaments are described by straight lines, $\kappa =0$, and thus $E_B=E_c= 0$. Thus, a general expression for the torque can be obtained using the chain rule $M= \frac{\partial E_{tot}}{\partial \theta} = \frac{\partial E_{tot}}{\partial \lambda} \frac{\partial \lambda}{\partial \theta}$, with $E_{tot} = E_t + E_s$. Then, 
$$\frac{M}{\mu V} = \left( 1- \frac{1}{\lambda^3} \left(1+\frac{1}{4}\left(\frac{R_0}{L_0} \right)^2 \right) \right) \frac{\partial \lambda^2}{\partial \theta} + \frac{1}{\lambda} \left(\frac{R_0}{L_0} \right)^2 \theta,$$
Then, using Eq.~\ref{eq:lambda1}, we have, in the Hookean limit
\bea
\frac{M}{\mu V} = \frac{3}{2}\left(\frac{W}{L_0}\right)^2\frac{\Delta L}{L_0} \sin \theta + \left( \frac{R_0}{L_0}\right)^2\theta.
\label{eq:MHookbeforecontact}
\eea
Thus, for $\theta < \theta_c$, the ratio between the stretch and twist contributions is given approximately by the stretch/twist ratio
$\mathcal{R} = 6 \frac{\Delta L}{L_0} \left(\frac{W}{D_0}\right)^2$.
For $W \approx D_0$, twist contribution dominates as $\mathcal{R} \approx 6 \Delta L/L_0 \ll 1$ in the Hookean limit and $\frac{M}{\mu V} \approx \left( \frac{R_0}{L_0}\right)^2\theta$ is pre-stretch independent. For large $W$ (i.e. $W\gg D_0$), the stretch/twist ratio $\mathcal{R}$ depends both on geometry and pre-stretch. When $\frac{\Delta L}{L_0} \ll \left(\frac{D_0}{W}\right)^2\ll 1$ (or, equivalently, $\mathcal{R} \ll 1$), the torque $\frac{M}{\mu V} \approx \left( \frac{R_0}{L_0}\right)^2\theta$ is still independent on the pre-stretch. But, for larger pre-stretch $\left(\frac{D_0}{W}\right)^2 \ll \frac{\Delta L}{L_0} \ll 1$ (or, equivalently, $\mathcal{R} \gg 1$), stretch contribution dominates yielding a torque that increases linearly with the pre-stretch and reads $\frac{M}{\mu V} = \frac{3}{2}\left(\frac{W}{L_0}\right)^2\frac{\Delta L}{L_0} \sin \theta$ . 

\subsubsection{After Contact}
We can solve the problem of finding the bundle length $\LH$ asymptotically after contact, and show that $\partial E_{tot} /\partial \LH = 0$ reduces to $\partial \lambda /\partial \LH = 0$. Consistent with numerical analysis which finds bending and contact contributions to be small, they are not included. We will check a posteriori that the bending contribution to the total energy is negligible in the limit $L_0\gg \theta R$.  Using Eq.~(\ref{eq:lambdahelix}), the minimization scheme yields after some algebra:
\bea
l_B=1-l_T = \frac{\dq}{\dw+\dq}.
\label{eq:LB}
\eea
Substituting Eq.~(\ref{eq:LB}) in Eq.~(\ref{eq:lambdahelix}), 
\bea
\lambda = \lambda_0\,\sqrt{1 + \left(\dw+\dq\right)^2}.
\label{eq:lambdaa}
\eea
Using Eq.~(\ref{eq:lambda}), Eq.~(\ref{eq:LBs}), and Eq.~(\ref{eq:LB}), we also have that $\cLH/\cL = \sqrt{l_B^2+\Delta q^2}/\sqrt{1+(\Delta q + \Delta w)^2} = l_B$. Likewise, we obtain $\cLT/\cL = l_T$. Note that $\theta_c$, $\dq$, and $dw$ should depend on $\lambda$ through $R$. To obtain a closed form of expression for $\lambda$, we take $R \approx R_0/\sqrt{\lambda_0}$. This gives an accurate expression if $\lambda$ remains close to $\lambda_0$, which occurs when $\Delta w$ and $\Delta q$ are small. 
Now, we evaluate the bending energy to check that it is indeed negligible in the limit $R\theta/L \ll 1$ and $W/L\ll 1$. Using Eqs~(\ref{eq:kappa}),  (\ref{eq:Eb}) and (\ref{eq:lambdaa}), we find that $\kappa R\approx (\dq + \dw)^2$ and $E_b \sim (\dq + \dw)^3\dq$. Interestingly, we find that the bending energy increases with $W$ because the curvature increase with $W$ overcompensates for the decrease in $\cLH/\cL = l_B$ (see Eq.~\ref{eq:LB}). Then, using Eqs.~(\ref{eq:Et}) and (\ref{eq:Es}), we find that $E_s \sim (\dq + \dw)^2$, $E_t \sim q_0^2$, and $E_b \sim (\dq + \dw)^3\dq \ll E_s, E_t$ when $\dq, \dw \ll 1$. For five turns and $W = 40\,$mm, $\dq\approx 1/3$ and $\dw \approx 1/5$ are not much smaller than unity which explains why bending energy is predicted to be a fraction of the twisting and stretching energy. Thus, bending energy can be neglected at a small enough twist angle. 

Finally, the torque after self-contact is given by substituting in Eq.~(\ref{eq:lambdaa}) in Eq.~(\ref{eq:Et}) and Eq.~(\ref{eq:Es}). Using the chain rule as before, we have in the Hookean limit
\bea
\frac{M}{\mu V}  = \left( \frac{R_0}{L_0} \right)\left(6\frac{\Delta L}{L_0} \dw + \q0\right).
\label{eq:torque_analin_aftercontact}
\eea
We compare these asymptotic calculations carried out in the case of Hookean-response with numerical calculations with the full elasto-geometric model in cases where the filaments are separate in Fig.~\ref{fig:linl}(b,c). One can note from Fig.~\ref{fig:linl}(b)  that the spacing between the filament $W>2R_0$ introduces an additional term in the torque, proportional to $\dw$. This term introduces an offset in the torque independent of $\theta$ and scales as $(RW/L_0^2)\,(\Delta L/L_0)$. Further comparing the two calculations for small $\Delta L/L_0$, we observe deviations for sufficiently large $W/L_0$ (see Fig.~\ref{fig:linl}(c)). This can be understood from the fact that the filaments are stretched significantly as $\theta$ is increased to $180^{\circ}$.

\subsection{Prescribed torque}

\begin{figure}
    \centering
    \includegraphics[width = 8cm]{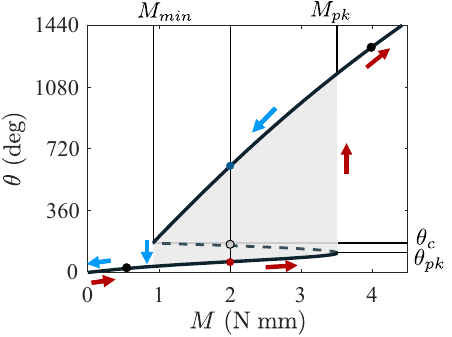}
    \caption{ The twist angle versus torque profile (solid and dashed black lines) under torque prescribed conditions using our elasto-geometric model with $\Delta L/L_0 = 1$\%, and $W/D_0 = 16$. Unlike for twist prescribed experiments, the profile is multi-valued for torque values between the minimum $M_{min}$ and peak value $M_{pk}$. The multi-valued profile leads to a hysteresis cycle (see gray area and arrows). Vertical lines at $M_{min}$ and $M_{pk}$ corresponds to dynamics jumps from one stable branch to the other. $\theta_{pk}$ is the twist angle corresponding to the peak value of the torque.}
    \label{fig:hysteresis}
\end{figure}
The analysis which enables us to determine  the torque required to twist the filament-pair as a function of angle also allows us to determine the angle reached if a prescribed torque is applied. In Fig.~\ref{fig:hysteresis}, we show the twist angle versus torque obtained from our elasto-geometric model ($\Delta L/L_0 = 1$\%, $W/D_0 = 16$) by inverting the relation $M(\theta)$ numerically. Unlike under the twist prescribed conditions, the profile is multi-valued. We predict two stable branches (solid black line) and one unstable branch (dashed black line) which corresponds to the decreasing part of the peak in twist angle controlled experiments. Outside this interval, there is only one stable branch. Thus, if the torque is increased quasi-statically from zero, the resulting twist angle increases slowly until $M=M_{pk}$ where the twist angle jumps dynamically to a larger equilibrium value on the stable top branch. Now, decreasing the torque starting from a value larger than $M_{pk}$, the twist angle decreases until $M=M_{min}$, where the twist angle jumps dynamically to the stable bottom branch. In a torque controlled experiment, we thus have 
a hysteresis cycle (gray area) which originates from the existence of a non-monotonous torque profile. 
The hysteretic behavior can be expected to vanish as $W \rightarrow D_0$, i.e. when the torque profile becomes monotonous.

We provide analytic expressions for the twist angle in torque controlled experiments. In case of filaments which are in contact, this corresponds to simply inverting the relation given by Eq.~(\ref{eq:M2R}), and rewriting $\theta$ in terms of $M$, i.e. 
\bea 
\theta =  \frac{1}{\mu V} \left( \frac{L_0}{R_0} \right)^2  \left(1 + 6 \frac{\Delta L}{L_0} \right)^{-1} {M}.
\eea

However, in the case where filaments are separate, there can be more than one stable angle depending on the applied torque in relation to the peak reached before the filaments come in contact, and the minimum when the filaments first come in contact. This can be easily seen by considering the intersection of horizontal lines corresponding the prescribed torque with $M$-$\theta$ plots shown in Fig.~\ref{fig:torque}(a,b).  

Before filaments come in contact, the stable angle can be obtained in general by solving Eq.~(\ref{eq:MHookbeforecontact}) numerically, but in the case of large pre-stretch (i.e.  $\left(\frac{D_0}{W}\right)^2 \ll \frac{\Delta L}{L_0} \ll 1$), we have, 
\bea
\theta = \sin^{-1}\left(\frac{2 }{3} \frac{M}{\mu V} \frac{L_0^3 }{\Delta L\, W^2 } \right).
\label{eq:theta_bef}
\eea
After contact, the twist angle that can be reached for a prescribed torque is obtained by inverting Eq.~(\ref{eq:torque_analin_aftercontact}), yielding
\bea
\theta =\frac{M}{\mu V} \left(\frac{L_0}{R_0}\right)^2- 6 \frac{\Delta L}{L_0}\frac{W}{R_0},
\label{eq:angle_analin_aftercontact}
\eea
where we use the fact that $\Delta w  \approx W/L_0$ for $W\gg D_0$. 

Thus, we find that two stable angles are possible for filaments where $W > D_0$ when applied torque is between the torque-minima when filaments come in contact and the peak value encountered before filaments come in contact. The angle reached in practice then depends on the initial conditions. In the case of filaments which are initially untwisted and parallel, $\theta$ corresponds to the angle before contact given by~Eq.~(\ref{eq:theta_bef}). However for initial twist angles corresponding to those greater than when filaments are contact, $\theta$ reached corresponds to Eq.~(\ref{eq:angle_analin_aftercontact}).

\section{Conclusions}
In summary, we investigated the torque required to twist a pair of extensible filaments that can be modeled as a neo-Hookean material. Geometry is shown to play an important role in determining the torque required to twist the filament pair.  By performing torque measurements and calculating them using an energy based approach, we have developed a deeper understanding of how energy is distributed in a twisted filament pair. We find that a surprisingly large amount of energy is stored in stretching. In fact, it is typically greater than in the twisting component for sufficiently large separation distances or pre-stretching. 

The overall torque profile, namely the rapid initial increase and peak, and then a slower increase is a feature of the geometry of the system. These features are true not only for nonlinear neo-Hookean materials, but for Hookean materials as well. We provide analytical formulas in the Hookean limit which can prove useful in calculating the energy in twisted filaments which are increasingly being considered in energy storage applications. These formulas can be inverted to obtain the twist angles reached for a given applied torque and initial twist angle. Further, our results further invite critical examination of analysis of twisted polymeric systems which focus on bending while considering the elastic energy stored in the system, and neglect stretching~\cite{Charvin2005,Panaitescu2018}.

\begin{center}
{\bf Acknowledgements}
    
\end{center}

This work was supported under U.S. National Science Foundation grant DMR-2005090.

\bibliographystyle{unsrt}

\begin{thebibliography}{10}

\bibitem{Bohr2011}
J.~{Bohr} and K.~{Olsen}.
\newblock {The ancient art of laying rope}.
\newblock {\em EPL (Europhysics Letters)}, 93(6):60004, March 2011.

\bibitem{legrain2016let}
Antoine Legrain, Erwin~JW Berenschot, Leon Abelmann, Jos{\'e} Bico, and Niels~R
  Tas.
\newblock Let's twist again: elasto-capillary assembly of parallel ribbons.
\newblock {\em Soft matter}, 12(34):7186--7194, 2016.

\bibitem{weiner2020mechanics}
Nicholas Weiner, Yashraj Bhosale, Mattia Gazzola, and Hunter King.
\newblock Mechanics of randomly packed filaments—the “bird nest” as
  meta-material.
\newblock {\em Journal of Applied Physics}, 127(5), 2020.

\bibitem{Slepukhin2021}
Valentin~M. Slepukhin, Maximilian~J. Grill, Qingda Hu, Elliot~L. Botvinick,
  Wolfgang~A. Wall, and Alex~J. Levine.
\newblock Topological defects produce kinks in biopolymer filament bundles.
\newblock {\em Proceedings of the National Academy of Sciences},
  118(15):e2024362118, 2021.

\bibitem{Gao2021}
Xiang Gao, Yifeng Hong, Fan Ye, James~T. Inman, and Michelle~D. Wang.
\newblock Torsional stiffness of extended and plectonemic dna.
\newblock {\em Phys. Rev. Lett.}, 127:028101, Jul 2021.

\bibitem{Plumb2022}
Thomas~B. Plumb-Reyes, Nicholas Charles, and L.~Mahadevan.
\newblock Combing a double helix.
\newblock {\em Soft Matter}, 18:2767--2775, 2022.

\bibitem{seguin2022twist}
Antoine Seguin and J{\'e}r{\^o}me Crassous.
\newblock Twist-controlled force amplification and spinning tension transition
  in yarn.
\newblock {\em Physical Review Letters}, 128(7):078002, 2022.

\bibitem{patil2023}
Vishal~P. Patil, Harry Tuazon, Emily Kaufman, Tuhin Chakrabortty, David Qin,
  Jörn Dunkel, and M.~Saad Bhamla.
\newblock Ultrafast reversible self-assembly of living tangled matter.
\newblock {\em Science}, 380(6643):392--398, 2023.

\bibitem{Grandgeorge2021}
Paul Grandgeorge, Changyeob Baek, Harmeet Singh, Paul Johanns, Tomohiko~G.
  Sano, Alastair Flynn, John~H. Maddocks, and Pedro~M. Reis.
\newblock Mechanics of two filaments in tight orthogonal contact.
\newblock {\em Proceedings of the National Academy of Sciences},
  118(15):e2021684118, 2021.

\bibitem{Moshe2004}
Moshe Shoham.
\newblock {Twisting Wire Actuator}.
\newblock {\em Journal of Mechanical Design}, 127(3):441--445, 07 2004.

\bibitem{Guzek2012}
James~J. Guzek, Conrad Petersen, Stephane Constantin, and Hod Lipson.
\newblock {Mini Twist: A Study of Long-Range Linear Drive by String Twisting}.
\newblock {\em Journal of Mechanisms and Robotics}, 4(1), 02 2012.
\newblock 014501.

\bibitem{kwon2014high}
C.~H. Kwon, S.~H. Lee, Y.-B. Choi, J.~A. Lee, S.~H. Kim, H.~H. Kim, G.~M.
  Spinks, G.~G. Wallace, M.~D. Lima, M.~E. Kozlov, R.~H. Baughman, and S.~J.
  Kim.
\newblock High-power biofuel cell textiles from woven biscrolled carbon
  nanotube yarns.
\newblock {\em Nat. Commun.}, 5(1):1--7, 2014.

\bibitem{Haines2016}
Carter~S. Haines, Na~Li, Geoffrey~M. Spinks, Ali~E. Aliev, Jiangtao Di, and
  Ray~H. Baughman.
\newblock New twist on artificial muscles.
\newblock {\em Proceedings of the National Academy of Sciences}, 113:11709 --
  11716, 2016.

\bibitem{charles2019topology}
Nicholas Charles, Mattia Gazzola, and L~Mahadevan.
\newblock Topology, geometry, and mechanics of strongly stretched and twisted
  filaments: solenoids, plectonemes, and artificial muscle fibers.
\newblock {\em Physical review letters}, 123(20):208003, 2019.

\bibitem{Thompson1996}
John Thompson and A.~Champneys.
\newblock From helix to localized writhing in the torsional post-buckling of
  elastic rods.
\newblock {\em Proceedings: Mathematical, Physical and Engineering Sciences},
  452:The Royal Society--, 01 1996.

\bibitem{Audoly2016}
Basile Audoly.
\newblock {\em CHAPTER 1: Introduction to the elasticity of rods}, pages 1--24.
\newblock 01 2016.

\bibitem{Olsen2012}
Kasper Olsen and Jakob Bohr.
\newblock Geometry of the toroidal n-helix: optimal-packing and zero-twist.
\newblock {\em New Journal of Physics}, 14(2):023063, feb 2012.

\bibitem{Grason2015}
Gregory~M. Grason.
\newblock Colloquium: Geometry and optimal packing of twisted columns and
  filaments.
\newblock {\em Rev. Mod. Phys.}, 87:401--419, May 2015.

\bibitem{Neukirch2002}
S.~Neukirch and G.~van~der Heijden.
\newblock Geometry and mechanics of uniform n-plies: From engineering ropes to
  biological filaments.
\newblock {\em Journal of Elasticity}, 69:41--72, 11 2002.

\bibitem{Grason2012}
Gregory~M. Grason.
\newblock Defects in crystalline packings of twisted filament bundles. i.
  continuum theory of disclinations.
\newblock {\em Phys. Rev. E}, 85:031603, Mar 2012.

\bibitem{Panaitescu2018}
Andreea Panaitescu, Gregory~M. Grason, and Arshad Kudrolli.
\newblock Persistence of perfect packing in twisted bundles of elastic
  filaments.
\newblock {\em Phys. Rev. Lett.}, 120:248002, Jun 2018.

\bibitem{Hanlan2023}
Jesse~M. Hanlan, Gabrielle~E. Davis, and Douglas~J. Durian.
\newblock Twist and measure: characterizing the effective radius of strings and
  bundles under twisting contraction.
\newblock {\em Soft Matter}, 19:4315--4322, 2023.

\bibitem{Charvin2005}
Gilles Charvin, Alexander Vologodskii, David Bensimon, and Vincent Croquette.
\newblock Braiding dna: Experiments, simulations, and models.
\newblock {\em Biophysical Journal}, 88:4124--36, 07 2005.

\bibitem{Bruss2013}
Isaac~R. Bruss and Gregory~M. Grason.
\newblock Topological defects{,} surface geometry and cohesive energy of
  twisted filament bundles.
\newblock {\em Soft Matter}, 9:8327--8345, 2013.

\bibitem{Chopin2022}
Julien Chopin and Arshad Kudrolli.
\newblock Tensional twist-folding of sheets into multilayered scrolled yarns.
\newblock {\em Science Advances}, 8(14), apr 2022.

\bibitem{Audoly2010}
Basile Audoly and Yves Pomeau.
\newblock {\em Elasticity and Geometry}.
\newblock Oxford University Press, New York, 2010.

\bibitem{Ogden1984}
Ray~W. Ogden.
\newblock {\em Non-Linear Elastic Deformations}.
\newblock Dover, New York, 1984.

\bibitem{o2006elementary}
Barrett O'neill.
\newblock {\em Elementary differential geometry}.
\newblock Elsevier, 2006.

\bibitem{rivlin1951large}
Ronald~S. Rivlin and D.W. Saunders.
\newblock Large elastic deformations of isotropic materials vii. experiments on
  the deformation of rubber.
\newblock {\em Philosophical Transactions of the Royal Society of London.
  Series A, Mathematical and Physical Sciences}, 243(865):251--288, 1951.

\bibitem{rivlin1948large}
R.S. Rivlin.
\newblock Large elastic deformations of isotropic materials. iii. some simple
  problems in cyclindrical polar co-ordinates.
\newblock {\em Philosophical Transactions of the Royal Society of London.
  Series A, Mathematical and Physical Sciences}, 240(823):509--525, 1948.

\bibitem{Ghatak2005}
A.~Ghatak and L.~Mahadevan.
\newblock Solenoids and plectonemes in stretched and twisted elastomeric
  filaments.
\newblock {\em Phys. Rev. Lett.}, 95:057801, Jul 2005.

\bibitem{johnson1987contact}
Kenneth~Langstreth Johnson.
\newblock {\em Contact mechanics}.
\newblock Cambridge university press, 1987.

\end{thebibliography}

\end{document}